# Joint Optimization of Traffic Signal Control and Vehicle Routing in Signalized Road Networks using Multi-Agent Deep Reinforcement Learning

Xianyue Peng[a,b,c,d], Hang Gao[e], Gengyue Han[a,c,d], Hao Wang[*,a,c,d] and Michael Zhang[b]

[a]Jiangsu Key Laboratory of Urban ITS, Southeast University, Nanjing, Jiangsu 211189, China; [b]the Department of Civil and Environmental Engineering, University of California, Davis, CA 95616, USA; [c]Jiangsu Province Collaborative Innovation Center of Modern Urban Traffic Technologies, Nanjing, Jiangsu 211189, China; [d]School of Transportation, Southeast University, Nanjing, Jiangsu 211189, China; [e]the Institute of Transportation Studies, University of California, Davis, CA 95616, USA.

## Abstract

Urban traffic congestion is a critical predicament that plagues modern road networks. To alleviate this issue and enhance traffic efficiency, traffic signal control and vehicle routing have proven to be effective measures. In this paper, we propose a joint optimization approach for traffic signal control and vehicle routing in signalized road networks. The objective is to enhance network performance by simultaneously controlling signal timings and route choices using Multi-Agent Deep Reinforcement Learning (MADRL). Signal control agents (SAs) are employed to establish signal timings at intersections, whereas vehicle routing agents (RAs) are responsible for selecting vehicle routes. By establishing relevance between agents and enabling them to share observations and rewards, interaction and cooperation among agents are fostered, which enhances individual training. The Multi-Agent Advantage Actor-Critic algorithm is used to handle multi-agent environments, and Deep Neural Network (DNN) structures are designed to facilitate the algorithm's convergence. Notably, our work is the first to utilize MADRL in determining the optimal joint policy for signal control and vehicle routing. Numerical experiments conducted on the modified Sioux network demonstrate that our integration of signal control and vehicle routing outperforms controlling signal timings or vehicles' routes alone in enhancing traffic efficiency.

**1. Introduction**

Traffic signal control and vehicle routing are recognized as effective measures to alleviate traffic congestion and enhance traffic efficiency in urban road networks. The signal settings are intrinsically linked to the route decisions made by drivers. This is because the traffic control system is designed to improve network performance, which is accomplished through a comprehensive analysis of the network flow patterns that encompasses drivers' route decisions. Conversely, particular signal controls can have an impact on the cost of different routes, prompting drivers to select particular routes for self-optimization or to achieve system optimization. The mutually beneficial objectives and interdependence of signal control and vehicle routing make collaborative optimization achievable, resulting in a substantial enhancement in network performance.

Connected and autonomous vehicles possess the capability to establish dependable two-way communications between vehicles and infrastructure, enabling the possibility of achieving a system optimum through synergistic vehicle routing and traffic signal control Li, Mirchandani, and Zhou (2015). Fig.1 illustrates a simplified network and the influence of a collaborative strategy involving signal control and vehicle routing. By rerouting the vehicles of $(O_1, D_1)$, the arrival times of conflicting vehicles in $S_1$ and $S_2$ are staggered, and optimized signal timings of $S_1$ and $S_2$ enables vehicles to leave the intersection as soon as possible.

Deep reinforcement learning (DRL) demonstrates exceptional potential in managing dynamic and complex environments, optimizing long-term performance, and making real-time decisions. As a result,

DRL has seen increasing applications in the field of intelligent transportation systems, including traffic-signal control, autonomous driving and traffic assignment(F. Yang et al. 2021a).

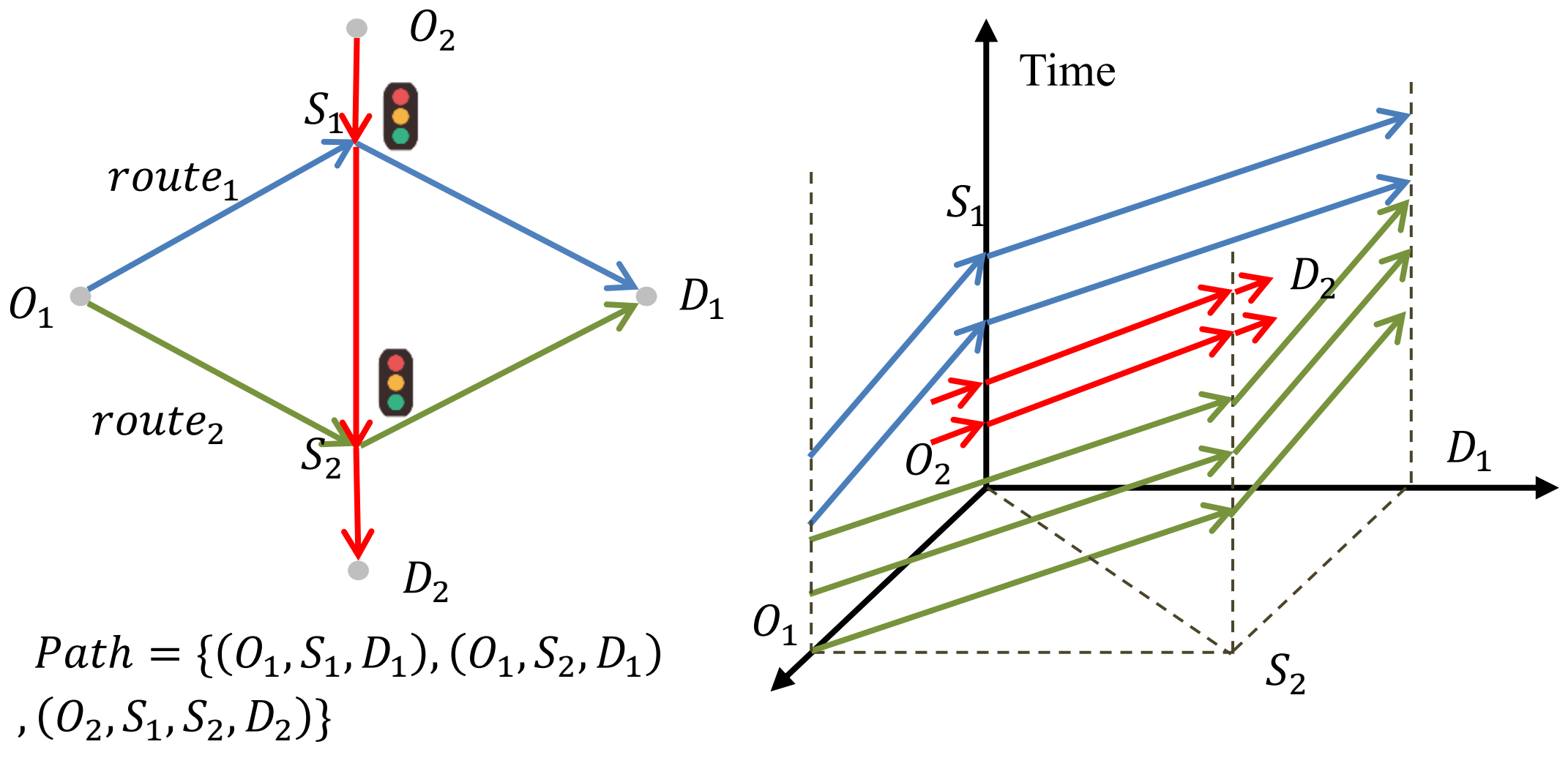

Figure 1 Simplified example for the joint signal control and vehicle routing

Multi-agent DRL (MADRL) is a technique that involves multiple agents learning in a shared environment(Grunitzki, Ramos, and Ieee 2020). Each agent takes actions to find the best rewards based on its observations, while the rewards and observations are influenced by the actions of other agents. This modeling approach is particularly suited for the joint signal control and vehicle routing problem, in which signalized intersections and connected vehicles take actions to achieve their goals and consider the behavior of others as the underlying dynamics of the environment. In other words, the MADRL approach facilitates coordination and collaboration between agents in the road network to find the optimal joint control policy.

This paper proposed a joint optimization of signal timings and vehicles' routes for signalized networks with time-varying demand. The optimal joint control policy is determined using a MADRL method.

The rest of the paper is organized as follows: Section 2 provides a review of the relevant literature in the field. Section 3 outlines the MADRL system's architecture, including the design of the signal controller and routing controller, which are critical components of the proposed system. Section 4 presents the Multi-Agent Advantage Actor-Critic DRL(MAA2C) algorithm used for addressing the joint control problem. Section 5 presents the results of numerical experiments and Section 6 concludes the paper.

## 2. Related works

### 2.1. Research on Combined Signal Control and Vehicle Routing Problem

Due to the interdependence between signal settings and drivers' behavior, the combined signal control and vehicle routing problem has interested researchers for a long time. R. E Allsop (1974) was the first to consider this interdependence, highlighting how signal settings impact the correlation between travel cost and traffic flow, and then affects route selection. He proposed a theoretical framework where the estimated results of traffic assignment (such as traffic flows) are treated as functions of the traffic control parameters. Gazis (1974) suggested that optimal control for congested traffic requires two operations: traffic allocation and signal settings. Several examples (Gazis 1974, D'Ans and Gazis 1976) demonstrated the impact of signal settings on queue size. Akcelik and Maher (1977) explored the interaction between route control and signal control, concluding that coupling signal control with route control is superior to signal control alone as it converts user equilibrium to system optimal.

Joint optimization of signal control and vehicle routing can be categorized into two main approaches: centralized control, which optimize the entire road network, and distributed control, which focus on optimizing local regions. Appendix Table A.1 provides an overview of the detailed information of related studies.

Centralized control aims to improve overall road network performance by optimizing signal timing and flow patterns throughout the network. Iterative approaches, such as those proposed by Gartner (1977), Wong and Yang (1999), Abdelfatah, Mahmassani, and Nrc (1998) and Ukkusuri, Doan, and Aziz (2013) were among the initial and widely used methods employed for this purpose. These iterative methods sequentially address the traffic assignment problem and signal control problem to obtain a mutually consistent solution(R. E. Allsop and Charlesworth 1977, Charlesworth 1977). Iterative optimization offers numerous practical benefits, as it leverages well-established and effective methods for signal timing and traffic assignment. These methods enable realistic road network representation, advanced travel cost estimation, and comprehensive signal timing plans. However, iterative algorithm may not always converge(C. Lee and Machemehl 2005) and can only obtain local optima, potentially resulting in suboptimal system performance (R. E. Allsop and Charlesworth 1977, Charlesworth 1977, Dickson 1981, Marcotte 1983).

In addition to iterative methods, researchers have explored various approaches for centralized control. These approaches include gradient-based optimization algorithms such as the gradient projection algorithm (Sheffi and Powell 1983) and the bundle-based gradient method (Chiou 2014); metaheuristic algorithms such as the genetic algorithm (Varia and Dhingra 2004, Yu, Ma, and Zhang 2018) and simulated annealing (Han et al. 2015); and decomposition optimization methods such as the Lagrangian decomposition algorithm (Li, Mirchandani, and Zhou 2015, Wang et al. 2020). Other details of these studies are presented in Appendix A. Local search algorithms and global optimization heuristic algorithms typically incorporate one or more traffic assignment processes in each iteration. For example, in the method proposed by Sheffi and Powell (1983), equilibrium flows are assigned in each step of the gradient update. Similarly, Yu, Ma, and Zhang (2018) formulated the dynamic user equilibrium control problem as a Mathematical Programs with Equilibrium Constraints (MPEC) problem and solved it using a GAMS solver in each generation of the genetic algorithm. The decomposition method proposed by Li, Mirchandani, and Zhou (2015), Wang et al. (2020), although it does not involve multiple instances of traffic assignment, still exhibits high computational complexity. This is because it considers the space-time trajectories of vehicles in the network and represents multiple intersection phases as MI-phases. With an increasing number of intersections, the number of MI-phases grows exponentially, leading to the "Curse of Dimensionality" in optimization. Due to the algorithm's complexity, although these methods provide dynamic modeling of traffic, they are challenging to achieve real-time control and apply to large-scale networks due to long computational times.

Distributed control methods optimize signal timing and vehicle routing in local areas, with a short optimization time, making them suitable for real-time control. Hawas (2000) developed an integrated real-time traffic assignment and signal control system. In this system, a group of local controllers is distributed throughout the network and cooperates to adjust traffic assignment and signal settings based on real-time traffic conditions. Le et al. (2017) proposed a utility-based optimization framework for distributed traffic control, employing the BackPressure principle to solve optimal turning fractions and signal plans. However, the vehicle routing in these methods is relies on local information, making it difficult to obtain a network-wide optimal routing scheme.

### 2.2. RL Applications in Signal Control and Vehicle Routing Problem

Traffic signal control and vehicle routing involve a series of decision-making processes that strive to accomplish a cumulative objective within a specified timeframe. To effectively tackle these sequential decision-making problems, reinforcement learning (RL) has emerged as a highly suitable technique (Qin et al. 2019), capable of considering both long-term and short-term performance. Consequently, RL has witnessed a growing number of applications in traffic control and vehicle routing.

Over the past few years, deep reinforcement learning (DRL) algorithms have been extensively employed in the studies of signal optimization. Liang et al. (2019) employed the DRL algorithm to control the traffic light cycle. To overcome the challenge of high dimensionality associated with large state and action spaces, J. Lee, Chung, and Sohn (2020) proposed a reinforcement learning algorithm that identifies the entire traffic state and simultaneously governs all traffic signals within the network. Given that networks with distributed signal controllers can be seen as multi-agent systems, several researchers have employed the MADRL algorithm to control signal timings at multiple intersections within these networks(Abdoos and Bazzan 2021). To tackle the multi-intersection signal control problem, S. Yang et al. (2021b) proposed a decentralized MADRL algorithm that employs an inductive heterogeneous graph neural network. Zhang et al. (2021) presented a cooperative multi-agent actor-critic DRL approach that utilizes an edge computing architecture. Their approach incorporates a cooperative mechanism that considers the contribution weight distributions of local agents to achieve global optimization in the context of traffic control. Abdoos and Bazzan (2021) developed a hierarchical multi-agent system to deal with large traffic signal control. This system comprises agents at different levels, with first-level agents leveraging reinforcement learning to identify the optimal policy and second-level agents employing a long short-term memory (LSTM) neural network to estimate the traffic state.

Numerous researchers have utilized RL to address vehicle routing problems(Mellouk, Hoceini, and Amirat 2007). In their recent study, Lazar et al. (2021) devised a routing strategy for autonomous vehicles (AVs) in mixed-autonomy traffic. Their approach involved employing a planner to govern the AVs and using DRL to learn the optimal policy. Others adopted a multi-agent structure, with each driver represented as an agent, to model the vehicle routing problem(Ramos, Bazzan, and da Silva 2018, Grunitzki, Ramos, and Ieee 2020). Ramos, Bazzan, and da Silva (2018) proposed a regret-minimizing method to tackle vehicle routing problem. In multi-agent RL, agents learn using regret as a reinforcement signal, which is calculated based on the agents' observed information and global information from the navigation app. Assuming that road users(agents) choose routes independently, Grunitzki, Ramos, and Ieee (2020) employed multi-agent RL to address traffic assignment problem, and designed three reward functions including expert-designed rewards, difference rewards and intrinsically motivated rewards.

While RL has gained popularity for optimizing signal control and vehicle routing with a consideration of both short-term and long-term impacts, the joint control problem using RL has received limited attention. Zhu et al. (2021) addressed this problem by jointly optimizing traffic signals and rerouting for connected and autonomous vehicles (CAVs) at a bottleneck intersection, but their approach did not utilize a multi-agent structure, potentially limiting scalability. Sun et al. (2023) developed a bi-directional hierarchical RL framework for coordinating traffic signal phases and AVs' turning directions. However, their focus on single intersection vehicle navigation could be improved by considering possible routes for the OD flows. They also mainly considered local traffic conditions around signals, in contrast to the real-world scenario where drivers choose routes based on path cost, including conditions on subsequent links.

### 2.3. Literature Summary and Paper's Contribution

The integration of signal control and vehicle routing has been supported by theoretical studies as a superior approach compared to relying solely on signal control. This integration can be approached through centralized control and distributed control methods. Existing centralized control methods aim to optimize network performance by optimizing signal timings and flow distribution from a network-wide perspective. However, these methods often face challenges such as high algorithm complexity, computational inefficiency, convergence to suboptimal solutions, and limited applicability to real-time control, especially for large-scale networks. On the other hand, distributed control methods focus on optimizing signal timing and vehicle routing in local areas, allowing for faster optimization and making them suitable for real-time control. However, distributed control relies on local information for vehicle routing, which makes it difficult to achieve a network-wide optimal routing scheme. The main challenge lies in effectively optimizing real-time control strategies that consider both the long-term and long-distance effects of route choice. This challenge arises due to the differing impacts of vehicle routing, which has a long-term and long-distance influence, and signal control, which primarily affects the short-term and local conditions. In summary, the integration of signal control and vehicle routing is a promising approach, but it requires tackling the difficulties of algorithm complexity, computational efficiency, and achieving network-wide optimization while considering both short-term and long-term effects of the control measures.

RL has become a popular approach for developing signal control and vehicle routing, as it optimizes a real-time control policy while considering both the short-term and long-term impacts of the policy. Despite this, there have been few studies using RL to tackle the joint control problem, and the absence of a multi-agent structure constrains the scalability of the solvable network. Furthermore, existing multi-agent RL methods for vehicle routing primarily consider individual vehicles as units, which may not be well-suited for integration with signal control agents.

The contribution of this paper is summarized as: 1) to the best of our knowledge, this is the first work that employs MADRL to find the optimal joint policy of signal control and vehicle routing; 2) a novel vehicle routing agent is designed to effectively address the multi-origin-destination routing problem, and it can be integrated into the same dimension as the signal control agents; 3) the interaction and cooperation among multiple agents are designed and corresponding DNN structures are constructed to facilitate convergence of the MADRL algorithm; 4) the performance of the control strategy under different compliance rates is discussed.

## 3 System architecture and controller design

### 3.1. System Architecture

The proposed method of joint optimization aims to improve the traffic efficiency in the road network by simultaneously controlling signal timings and vehicles' routes. The joint control problem can be formulated as a multi-agent structure, where Signal Control Agents (SAs) are responsible for adjusting signal timings at intersections, while Vehicle Routing Agents (RAs) determine the routes for the vehicles.

As shown in Fig.2, SAs are distributed in the network and each agent controls signal plan of each signalized intersection. At the same time, each RA is situated at the upstream edge of the fork point of the feasible routes within an Origin-Destination (OD) pair. It aims to help vehicles choose subsequent routes of vehicles for this OD pair on this edge. For instance, OD pair (O1, D1 exists three feasible routes, i.e., Route1, Route2, Route3, with two RAs positioned at the edge (O1, I1) and edge (I1, I2), where RA1 determines to choose Routes from 1,2, or 3; and RA2 decides to choose Routes from 1 or 2. The number of RAs depends on the number of feasible routes of OD pairs. In real-world scenarios, it will not be overly abundant, as several routs with high travel costs may be discarded in advance.

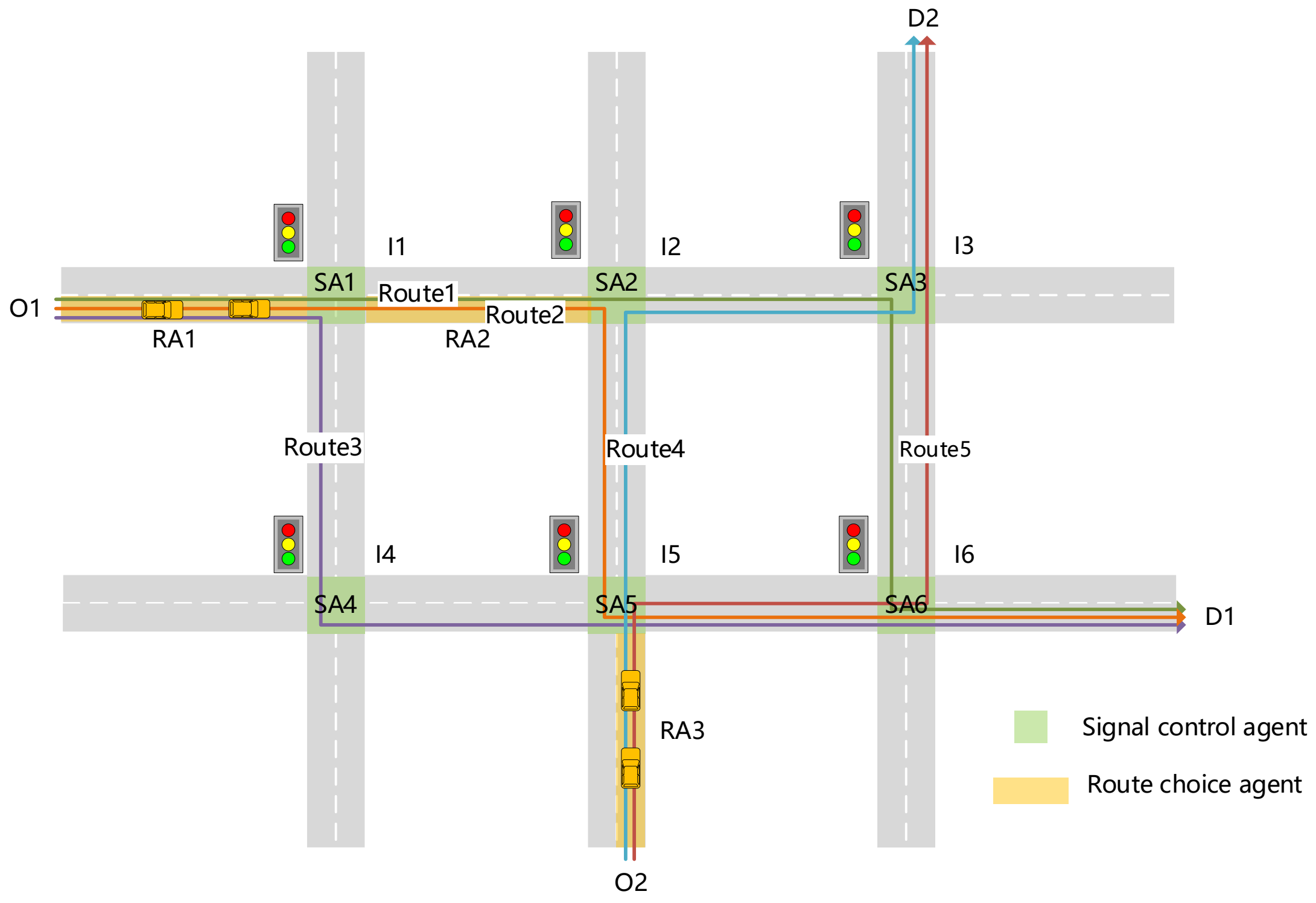

Figure 2 Example network

In MADRL, agents interact with the environment through a series of discrete control steps. The decision-making process of an agent can be described as a Markov decision process (MDP). At each control step $t$, given a state $s$, the agent takes an action $a$ according to its policy $\pi$. Then the environment returns a reward and subsequently updates the state. In DRL, the agent searches for the optimal policy within a continuous time series, whereby the optimal policy typically corresponds to the maximum cumulative reward over the long term.

Fig.3 shows the implementation architecture of the joint optimization method based on MAA2C

DRL approach. A decentralized training and execution approach is employed, whereby each agent has its own policy network and value network. This means that each agent to learn its own decision-making policy and value function independently, based on its own observations (referred to as local state) and rewards. As the actions of each agent can impact the next state of the environment, it can subsequently affect the behaviors of other agents within the system. To promote coordination among the agents, shared observations and rewards are utilized, enabling agents to exchange information and synchronize their behaviors. Additional details regarding this approach will be provided in Section 4.1.

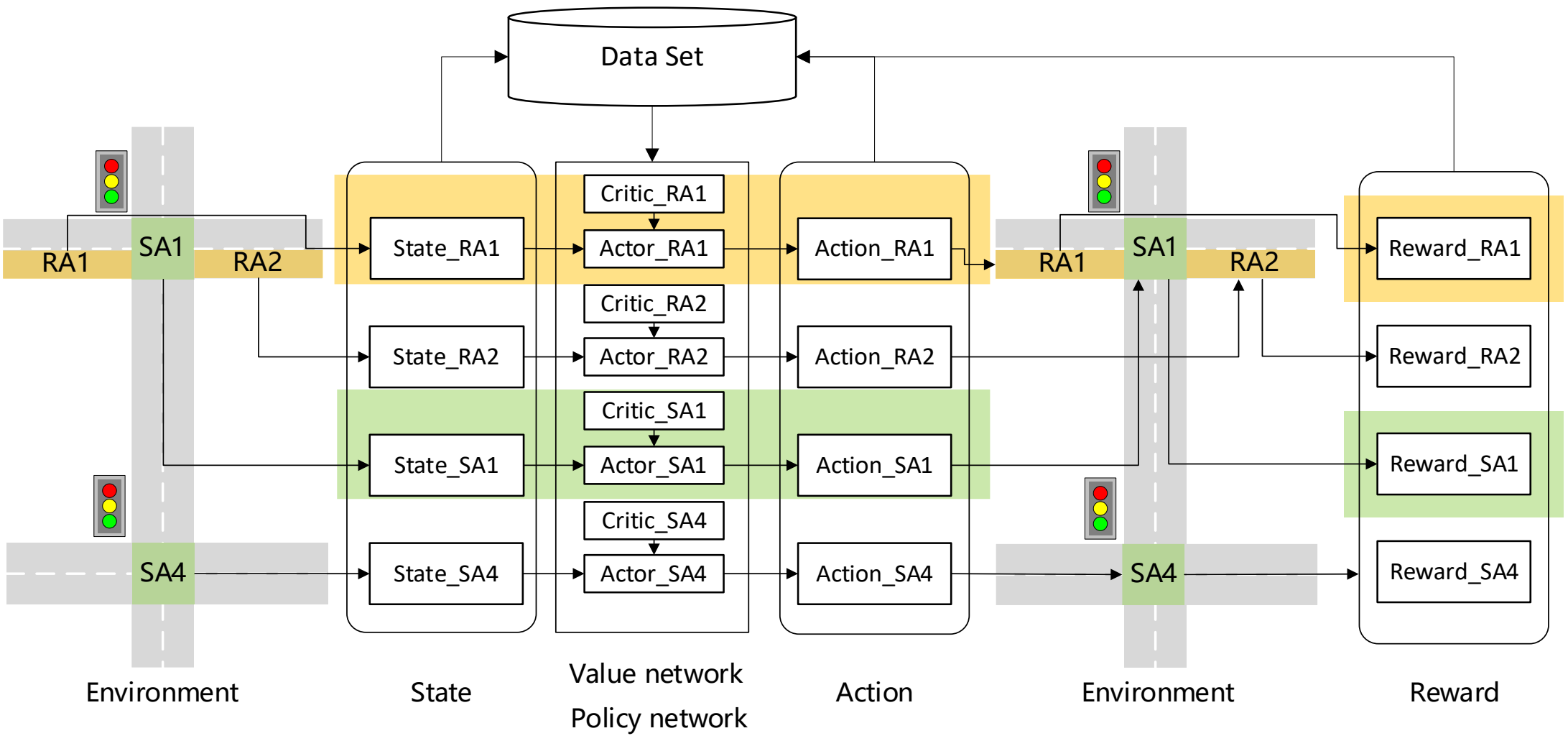

Figure 3 The implementation architecture of joint control

### 3.2. Signal Control Agent

SA optimizes the signal timings of an intersection, thereby reducing vehicle delay. The local state, action and reward definitions for SA are defined as follows.

#### 3.2.1 State representation

We consider the state of SA as a collection of indices for each lane approaching intersection. This collection is composed of the number of waiting vehicles, number of approaching vehicles, and the total waiting time. Specifically, the local state of SA is defined as

$$s_{i,t}^{\text{local}} = [s_{i,t}^{\text{wave}}, s_{i,t}^{\text{wait}}], \forall i \in C_S \quad (1.)$$

Where $C_S$ is the set of SAs, $s_{i,t}^{\text{local}}$ is the local state for agent $i$ at control step $t$. $s_{i,t}^{\text{wave}} = [s_{i,t,l}^{\text{wave}}]_{l \in L_i}$, $s_{i,t}^{\text{wait}} = [s_{i,t,l}^{\text{wait}}]_{l \in L_i}$. $s_{i,t,l}^{\text{wave}}$ is the number of waiting and approaching vehicles of approach lane $l$ for agent $i$ at control step $t$, $s_{i,t,l}^{\text{wait}}$ represents the cumulative waiting time of all vehicles in approach lane $l$ for agent $i$ within control step $t$, excluding waiting times that occur outside this control step. $L_i$ denotes the set of the approach lanes of intersection $i$.

### 3.2.2 Action definition

During each control step, we select one phase from the feasible phases to be activated as the action.

$$a_{i,t} = a_{i,t}^{p}, a_{i,t}^{p} \in Phase_i, \forall i \in C_S \quad (2.)$$

Where $Phase_i$ is the set of feasible phases for intersection $i$; $a_{i,t}^{p}$ is the selected phase of SA $i$ at control step $t$. $Phase_i$ are predetermined according to the layout of the intersection, and do not include transitional phases such as yellow. This is because if the chosen phase $a_{i,t}^{p}$ is identical to $a_{i,t-1}^{p}$, namely the activated phase at last control step $t$-1, the chosen phase should be executed immediately. Otherwise, the transition phase between $a_{i,t-1}^{p}$ and $a_{i,t}^{p}$ is executed. For example, if the signal for approach lane $l$ changes from green to red, then the transition phase of yellow is executed for that approach lane; Similarly, if the signal for approach lane $l$ changes from red to green, then transition phase of red will be executed.

### 3.2.3 Reward definition

To optimize signal control effectively, it is crucial to use a reward that is spatially decomposable and frequently measurable, which enables targeted analysis and adjustments to address congested lanes and allows for real-time adaptations to changing traffic conditions. Hence, we define the reward of SA as the total waiting time during the control step and the waiting time of the first vehicle since it stops. The second term is included to prevent a situation where a certain phase is not executed for a prolonged period.

$$r_{i,t}^{\text{local}} = -\sum_{l\in L_i}(r_{i,t,l}^{\text{wait}} + \alpha_1 r_{i,t,l}^{\text{fwait}}), \forall i \in C_S \quad (3.)$$

Where $r_{i,t,l}^{\text{wait}}$ is the total waiting time of vehicles in the approach lane $l$ for agent $i$ during control step $t$; $r_{i,t,l}^{\text{fwait}}$ represents the waiting time of the first vehicle in the queue since it stops. The coefficient $\alpha_1$ is used to adjust the impact of the reward $r_{i,t,l}^{\text{fwait}}$ on the overall reward calculation.

### 3.3. Vehicle Routing Agent

RA is located at the upstream road of the nodes where the target vehicles need to make a route choice and controls the subsequent routes of these vehicles. Fig.4 illustrate the related parameters and definitions for RA2 in the example network. The local state, action and reward of RAs are defined as follows.

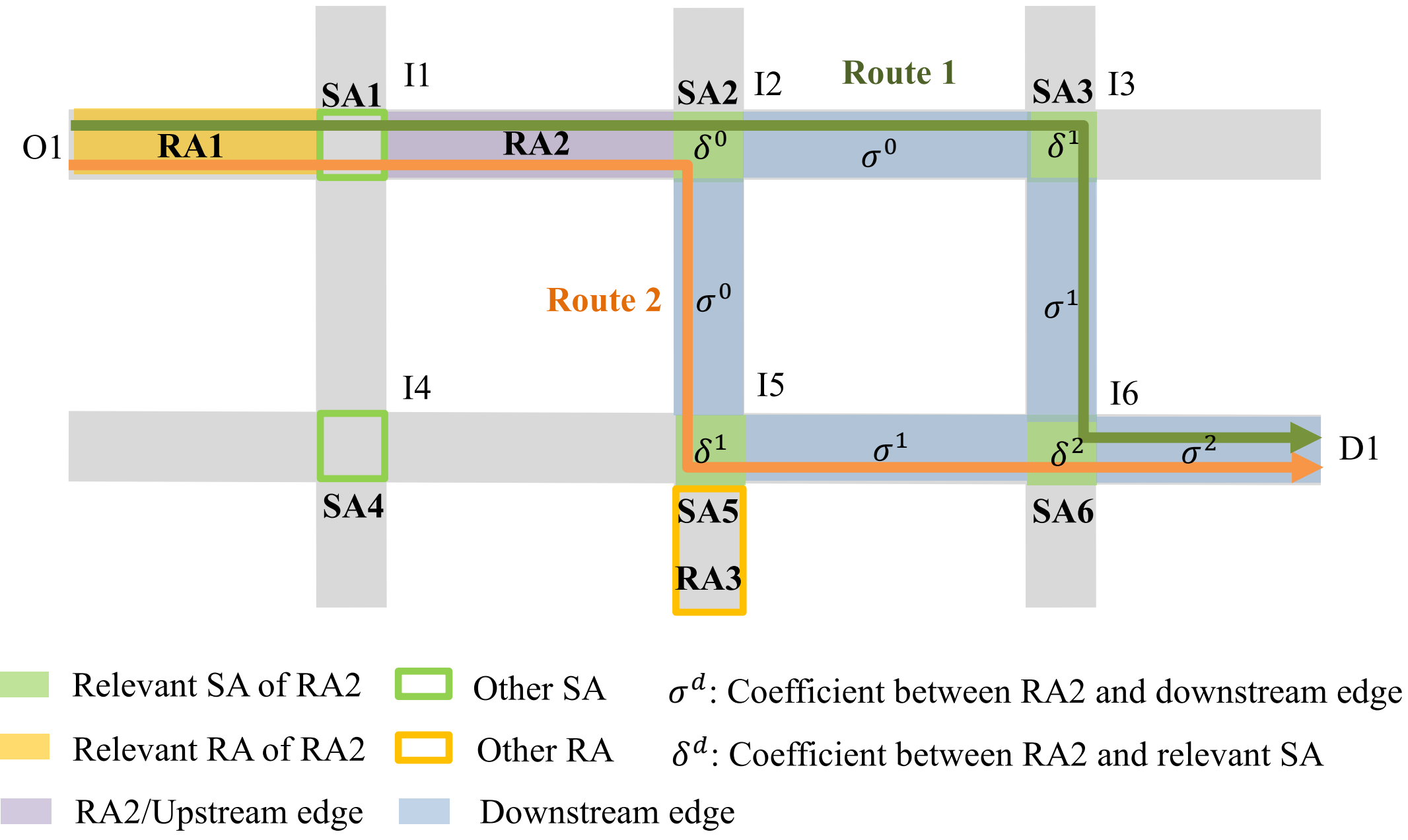

Figure 4 Related parameters and definitions for RA2 in the example network

#### 3.3.1. State representation

To make an appropriate route decision, we need to consider both the number of vehicles that require a route decision and the traffic conditions of the potential routes. As shown in Fig.4, the upstream edge leads to the path fork point, while the downstream edge is located on the route after the path fork point. The local state is defined as

$$s_{i,t}^{\text{local}} = [s_{i,t}^{\text{arrival}}, s_{i,t}^{\text{down}}], \forall i \in C_R \tag{4.}$$

Where $C_R$ is the set of RAs, $s_{i,t}^{\text{arrival}} = [s_{i,t,e}^{\text{arrival}}]_{e \in U_i}$, $s_{i,t,e}^{\text{down}} = [\alpha_e s_{i,t,e}^{\text{down}}]_{e \in D_i}$. $U_i$ denotes the upstream edge of RA $i$, and $D_i$ denotes the downstream edges of RA $i$. $s_{i,t,e}^{\text{arrival}}$ is the number of agent $i$'s vehicles on the upstream edge $e$, which need to decide following routes; $s_{i,t,e}^{\text{down}}$ is the total number of vehicles on the downstream edge $e$; $\alpha_{i,e}$ is the coefficient of downstream edge $e$, which is determined based on the location of the edge. Usually, the coefficient is smaller for edges that are farther away from the path fork point, indicating that their impact on the overall reward calculation is reduced in comparison to nearby edges. Specifically, we set $\alpha_{i,e}$: $\alpha_{i,e} = \sigma^{d_{i,e}}$, $\sigma$ is a parameter between 0 and 1, $d_{i,e}$ denotes the distance between agent $i$ and downstream edge $e$. As an example in Fig.4, $\alpha_{\text{RA2,(I2,I3)}} = \sigma^0, \alpha_{\text{RA2,(I2,I5)}} = \sigma^0, \alpha_{\text{RA2,(I3,I6)}} = \sigma^1, \alpha_{\text{RA2,(I5,I6)}} = \sigma^1, \alpha_{\text{RA2,(I6,D1)}} = \sigma^2$.

### 3.3.2. Action definition

At control step $t$, RA $i$ selects a route for vehicles positioned upstream of the route divergence points corresponding to the OD pair. RA $i$ 's action space consists of the available route options for these vehicles. In the specific example network, RA 2 has two route options: Route1 and Route2. This means RA 2 can choose between Route1 and Route2 for vehicles positioned upstream of divergence points at control step $t$.

Due to the edge capacity, the number of these vehicles during one control step is usually not large, and some of them may still be on the edge after this control step, which will follow the next action at the subsequent control step. Therefore, it is reasonable to choose the same route for vehicles during one control step.

$$a_{i,t} = a_{i,t}^{r}, a_{i,t}^{r} \in Route_i, \forall i \in C_R \tag{5.}$$

Where $Route_i$ is the set of feasible routes of RA $i$; $a_{i,t}^{r}$ is the selected route of RA $i$ at control step $t$.

### 3.3.3. Reward definition

The control objective of RA is to improve trip completion flow. Hence, the local reward is defined as

$$r_{i,t}^{\text{local}} = \alpha_2 r_{i,t}^{\text{arrived}}, \forall i \in C_R \tag{6.}$$

Where $r_{i,t}^{\text{arrived}}$ is the number of arrived vehicles of the origin and destination corresponding to RA $i$ during control step $t$. The coefficient $\alpha_2$ is used to adjust the impact of the reward $r_{i,t}^{\text{arrived}}$ on the overall reward calculation.

## 4. MAA2C DRL Algorithm for Joint Signal Control and Route Guidance

The MAA2C DRL algorithm is employed to address the proposed joint optimization problem, making use of deep neural networks (DNNs) for approximating both policy and value functions. The MAA2C algorithm is well-suited for scenarios involving multiple agents, where agents interact, cooperate, or compete. It facilitates distributed decision-making and rewards sharing among agents, offering interpretability and parallelism to accelerate learning. Furthermore, MAA2C adapts well to changing state spaces and uncertainties, making it particularly suited for complex, dynamic environments such as multi-robot systems, multi-agent games, or collaborative tasks like traffic management.

This section outlines the interaction among multiple agents in the proposed system architecture, the MAA2C algorithm, and the DNN configuration.

### 4.1. Interaction Between Multiple Agents

The proposed method considers the interaction among multiple agents. We define relevant agents to distinguish the degree of correlation among agents in the network. These relevant agents share observations and rewards, facilitating their individual training process.

The relevant RAs for an RA are the agents associated with the same OD pairs, and the relevant SAs for an SA are the agents located at its adjacent intersections. Notably, the relevant SAs for an RA are situated at intersections along the potential routes, while the relevant RAs for an SA are agents whose potential routes pass through its intersection. The relationship between SAs and RAs is mutual, meaning if an RA is associated with an SA, then the SA is also associated with the RA. Table 1 demonstrates the relevant agents of RAs and SAs in the example network as shown in Fig. 4.

Table 1 The definition of relevant agents

| Agent | Relevant SA | Relevant RA |
|---|---|---|
| **SA** | Agents at its adjacent intersection | Agents whose potential routes pass through its intersection |
| SA1 | SA2, SA4 | RA1 |
| SA2 | SA1, SA3, SA5 | RA1, RA2, RA3 |
| … | … | … |
| **RA** | Agents at intersections along the potential routes | Agents of the same OD pairs |
| RA1 | SA1, SA2, SA3, SA4, SA5, SA6 | RA2 |
| RA2 | SA2, SA3, SA5, SA6 | RA1 |
| RA3 | SA2, SA3, SA5, SA6 | - |

To enhance the connections between an agent and its relevant agents, we incorporate the shared state and fingerprint of the relevant agents as a part of the agent's own state. The fingerprints denote the relevant agents' policy at the previous control step. Given that the proposed control strategy is real-time and the duration of control step is relatively small, the current step policy is similar to that of last step (Chu et al. 2020). The state of SA and RA can be represented as

$$s_{i,t} = [s_{i,t}^{\text{local}}] \cup [s_{j,t}^{\text{wave}}]_{j \in C_i^{SS}} \cup [s_{j,t}^{\text{arrival}}]_{j \in C_i^{SR}} \cup [\pi_{j,t-1}]_{j \in C_i^{SS} \cup C_i^{SR}}, \forall\, i \in C_S \quad (7.)$$

$$s_{i,t} = [s_{i,t}^{\text{local}}] \cup [s_{j,t}^{\text{arrival}}]_{j \in C_i^{RR}} \cup [s_{j,t}^{\text{wave}}]_{j \in C_i^{RS}} \cup [\pi_{j,t-1}]_{j \in C_i^{RR} \cup C_i^{RS}}, \forall\, i \in C_R \quad (8.)$$

Where $C_i^{SS}$ is the set of relevant SAs of SA $i$, $C_i^{SR}$ is the set of relevant RAs of SA $i$, $C_i^{RR}$ is the set of relevant RAs of RA $i$, $C_i^{RS}$is the set of relevant SAs of RA $i$. Since $s_{j,t}^{\text{wait}}$ and $s_{j,t}^{\text{down}}$ of the relevant agent $j$ are not as strongly associated with agent $i$ as $s_{j,t}^{\text{wave}}$ and $s_{j,t}^{\text{arrival}}$, they are not considered in this situation.

Also, we take relevant agents' reward into consideration. The reward of SA and RA are calculated as follows,

$$r_{i,t} = r_{i,t}^{\text{local}} + \beta^{SS} \sum_{j \in C_i^{SS}} r_{j,t}^{\text{local}} + \sum_{j \in C_i^{SR}} \beta_{(i,j)}^{SR} r_{j,t}^{\text{local}}, \forall\, i \in C_S \quad (9.)$$

$$r_{i,t} = r_{i,t}^{\text{local}} + \sum_{j \in C_i^{RS}} \beta_{(i,j)}^{RS} r_{j,t}^{\text{local}}, \forall\, i \in C_R \quad (10.)$$

By definition, the local reward of RA $i$ equals to that of its relevant RA $j$, therefore, this term is not reiterated in Eq. (10). $\beta^{SS}$ is the discount parameters of relevant SA of SA; $\beta_{(i,j)}^{SR}$ is the discount parameters of relevant RA $j$ of SA $i$; $\beta_{(i,j)}^{RS}$ is the discount parameters of relevant SA $j$ of RA $i$.

Typically, the discount parameter is smaller for SA that are farther away from the RA. To be specific,

$$\beta_{(i,j)}^{\text{RS}} = \beta^{\text{RS}} \alpha_{i,j}, \forall\, i \in C_R, j \in C_i^{RS} \quad (11.)$$

$$\beta_{(i,j)}^{\text{SR}} = \beta^{\text{SR}} \alpha_{i,j}, \forall\, i \in C_S, j \in C_i^{SR} \quad (12.)$$

where $\beta^{\mathrm{RS}}, \beta^{\mathrm{SR}}$ are parameters between 0 and 1, and $\alpha_{i,j}$ is the coefficient between agent $i$ and agent $j$, which is determined based on the distance between these agents. Specifically, $\alpha_{i,j} = \delta^{d_{i,j}}$, $\delta$ is a parameter between 0 and 1, $d_{i,j}$ indicates the distance between agent $i$ and agent $j$. For example, in Fig2, $\alpha_{\mathrm{RA2,SA2}}=\delta^0$, $\alpha_{\mathrm{RA2,SA3}}=\delta^1$, $\alpha_{\mathrm{RA2,SA5}}=\delta^1$, $\alpha_{\mathrm{RA2,SA6}}=\delta^2$ and $\alpha_{\mathrm{SA2,RA2}}=\delta^0$, $\alpha_{\mathrm{SA3,RA2}}=\delta^1$, $\alpha_{\mathrm{SA5,RA2}}=\delta^1$, $\alpha_{\mathrm{SA6,RA2}}=\delta^2$.

### 4.2. MAA2C DRL Algorithm for Joint Control

#### 4.2.1. MAA2C algorithm

The MAA2C algorithm is a variant of the Advantage Actor-Critic (A2C) algorithm to handle multi-agent environments, where multiple agents coexist and interact with each other.

Each agent has its own decision-making policy $\pi_{\theta_i}(a_{i,t}|s_{i,t})$ and value function $V_{\omega_i}(s_{i,t})$. The Temporal Difference (TD) target of agent $i$ at control step $t$ can be calculated as

$$R_{i,t} = \sum_{\tau=t}^{t_B-1} \gamma^{\tau-t} r_{i,t} + \gamma^{t_B-t} V_{\omega_i}(s_{i,t_B}) \tag{13.}$$

Where $B$ is the transition batch, $B = \{s_t, a_t, r_t, s_{t+1}\}$, $t_B$ is the final control steps of the transition batch $B$; $\gamma$ is the discount factor, a value between 0 and 1. It determines the importance of future rewards in reinforcement learning. By balancing short-term gains and long-term rewards, the discount factor allows the agent to make optimal decisions over time.

---

**Algorithm** Multi-agent advantage actor critic algorithm

---

**Input:**

Initialize policy parameters $\theta_i$, value-function parameters $\omega_i$, empty transition buffer $D$

Initialize environment state, fingerprints, $t \leftarrow 1$

---

**While True:**

**For** $n$ steps:

**For** $i \in C_S \cup C_R$

Select action $a_{i,t}$ according to policy $\pi_{\theta_i}\left(a_{i,t} \middle| s_{i,t}\right)$

Get value $V_{\omega_i}(s_{i,t})$ according to value function $V_{\omega_i}(s_{i,t})$

Execute $a_t$ in the environment

Get next state $s_{t+1}$, reward $r_t$

Store $(s_t, a_t, r_t, V_{\omega_i}(s_{i,t}))$ in the transitions batch $B$

Update state $s_t \leftarrow s_{t+1}$, $t \leftarrow t + 1$

**For** $i \in C_S \cup C_R$:

Given $B = \{s_t, a_t, r_t, V_{\omega_i}(s_{i,t})\}$

Compute $R_t, A_t$

Update policy by one step of gradient descent using

$$\theta_i \leftarrow \theta_i + \tau_\theta \nabla_{\theta_i} J(\theta_i)$$

Update V-functions by one step of gradient descent using

$$\omega_i \leftarrow \omega_i - \tau_\omega \nabla_{\omega_i} L(\omega_i)$$

Empty the transitions batch $B$

**If** the simulation step size reaches the last step:

Reset environment state, fingerprints, $t \leftarrow 1$

**If** stopping condition reached:

**Return** policy parameters $\theta_i$, value-function parameters $\omega_i$

The advantage function $A_{i,t}$ can be calculated by

$$A_{i,t} = R_{i,t} - V_{\omega_i}(s_{i,t}) \quad (14.)$$

The value loss is as follows:

$$L(\omega_i) = \frac{1}{2n}\sum_{t\in T_B}(R_{i,t} - V_{\omega_i}(s_{i,t}))^2 \quad (15.)$$

Where $n$ is the number of steps in a batch; $T_B$ is the set of the control steps in the transition batch.

The policy gradient is as follows,

$$\nabla_{\theta_i} J(\theta_i) = \nabla_{\theta_i}\frac{1}{n}\sum_{t\in T_B}\left(log\pi_{\theta_i}(a_{i,t}|s_{i,t})A_{i,t} + \beta_{\text{en}}H\left(\pi_{\theta_i}(s_{i,t})\right)\right) \quad (16.)$$

The entropy of the policy $\pi_{\theta_i}$ is incorporated into the function of improved exploration to discourage premature convergence to suboptimal deterministic policies. This regularization term encourages a distribution with a "flat entropy" shape, which makes it less likely to converge on a single action during training(Williams and Peng 1991). By adding this entropy term to the objective function, the exploration of different actions is promoted, preventing the policy from getting stuck in local optima and promoting a more diverse exploration of the action space. $H\left(\pi_{\theta_i}(s_{i,t})\right)$ is the entropy, $H\left(\pi_{\theta_i}(s_{i,t})\right) = -\sum_{a_{i,t}\in V_{i,t}}\pi_{\theta_i}(a_{i,t}|s_{i,t})log\pi_{\theta_i}(a_{i,t}|s_{i,t})$, $V_{i,t}$ is the set of $a_{i,t}$, hyperparameter $\beta_{\text{en}}$ controls the strength of the entropy regularization term.

The MAA2C algorithm is shown as follows, where $s_t = \{s_{i,t}|i \in C_S \cup C_R\}$, $a_t = \{a_{i,t}|i \in C_S \cup C_R\}$, $r_t = \{r_{i,t}|i \in C_S \cup C_R\}$, $R_t = \{R_{i,t}|i \in C_S \cup C_R\}$, $A_t = \{A_{i,t}|i \in C_S \cup C_R\}$. Each batch has n steps. At each step, the policy and value networks are used to obtain the action $a_{i,t}$ and the value function $V_{\omega_i}(s_{i,t})$ for each agent $i$. These actions are then executed in the environment, resulting in a new state $s_{t+1}$ and reward $r_t$. The transition for each step is stored in a transition batch. After collecting a batch of data, the TD target and advantage for each agent can be calculated, and the policy

and value networks can be updated accordingly. If the simulation step size reaches the last step, the environment is reset, and the training continues with a new episode. The algorithm continues to collect data and update the networks until a stopping condition is met, such as reaching a maximum number of training steps or achieving a certain level of performance. Once the stopping condition is met, the policy parameters $\theta_i$, value-function parameters $\omega_i$ are returned.

### 4.2.2. DNN configuration

In this multi-agent algorithm, each agent has its own policy network and value network. The adopted DNNs consists of fully connected (FC) layers, long-short term memory (LSTM) layer (Hochreiter and Schmidhuber 1997) and policy /value FC layers. The policy and value DNN are trained separately, instead of sharing lower layers between them. Fig.5 presents the DNN structure of SA.

To avoid the confusion of state input, each type of state is processed by an independent FC layer(Chu et al. 2020). For SA, the states include wave states of itself and relevant SAs, wait states of itself, arrival states of relevant route agents and policy of relevant agents. For RA, the states include arrival states of itself and relevant RAs, downstream state if itself, wave states of relevant SAs and policy of relevant agents. The corresponding FC layers are constructed for these four type states.

Then all hidden data is merged and input into a LSTM layer. As for signal control and vehicle route problem, since network traffic flow is complex spatial-temporal data, the collected data of current states cannot reflect the complete traffic characteristics required for control strategy. Taking LSTM as a hidden layer can tackle this problem since it utilizes the features of time series within a longer time window. In LSTM network, previous cell states are conveyed, and each LSTM cell includes input gate, forget gate and output gate, which enable LSTMs to cope with sequential data with long-term dependencies.

For policy DNN, the output layer is a FC layer with the number of neurons in the FC layer matching the size of the action space. A softmax function is adopted to generate the policy, with each output neuron representing the probability of a specific action. For value DNN, the output is in the form of a single value, which represents the estimated value of the corresponding state.

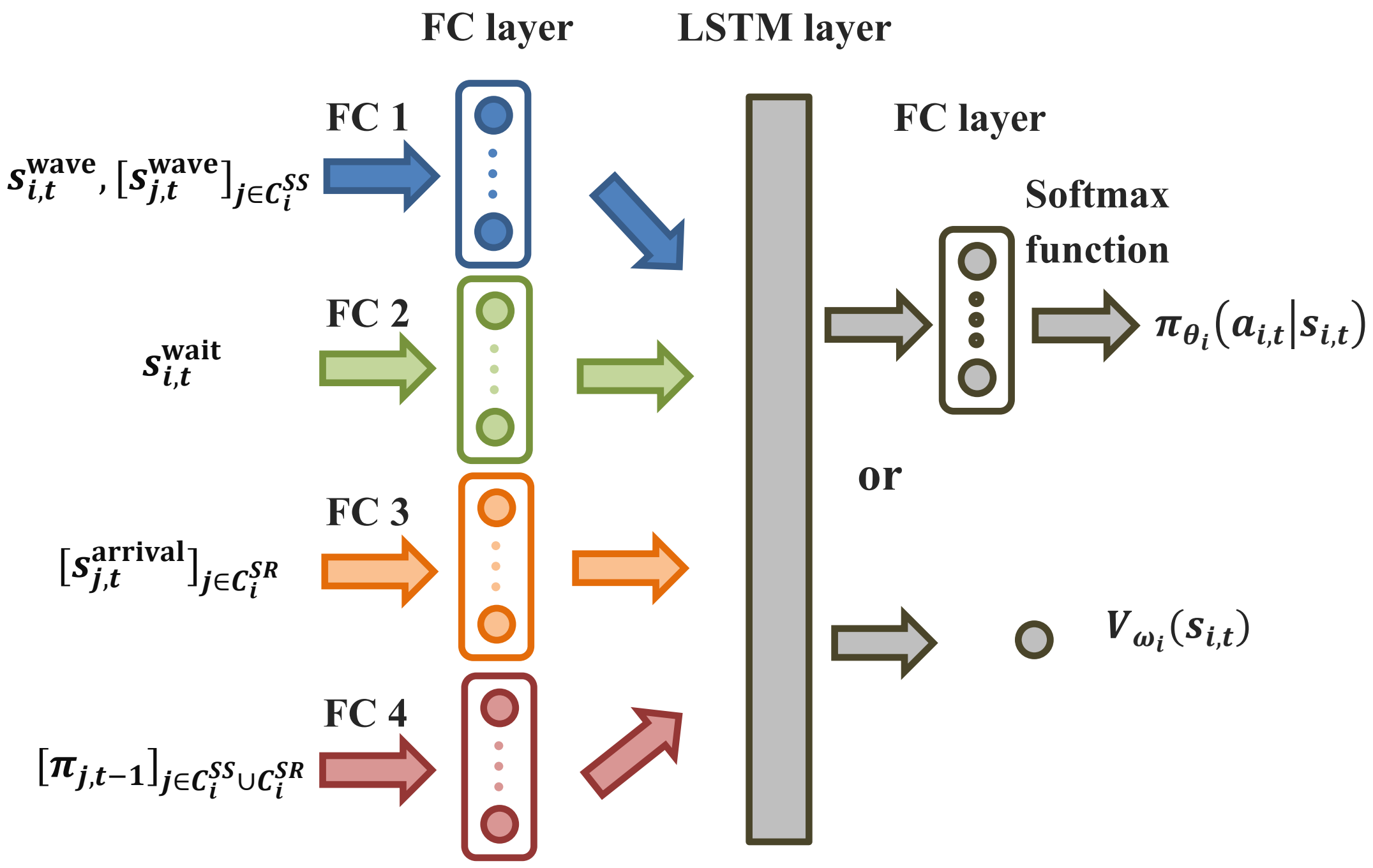

Figure 5 Policy/value DNN structure of SA

A gradient-based stochastic optimization method called adaptive moment estimation algorithm (Adam) (Kingma and Ba 2014) is adopted for DNN training. Adam is a prevailing optimizer for deep learning, which combines the advantages of AdaGrad(Duchi, Hazan, and Singer 2011) and RMSProp.

## 5. Numerical Experiments

### 5.1. Parameter Settings

#### 5.1.1. Environment settings

SUMO-simulated traffic environment is adopted to evaluate the effectiveness of the proposed method. We test the proposed joint control method in a modified Sioux Falls network. The simulated traffic environment for each episode lasts 3600 seconds, with the control step length and the duration of the transition phase set to 5 seconds.

Fig. 6 shows the layout of the network, which includes 17 signalized intersections， 7 intersections under priority control and 7 external intersections that serve as origin and destination nodes for traffic demand.

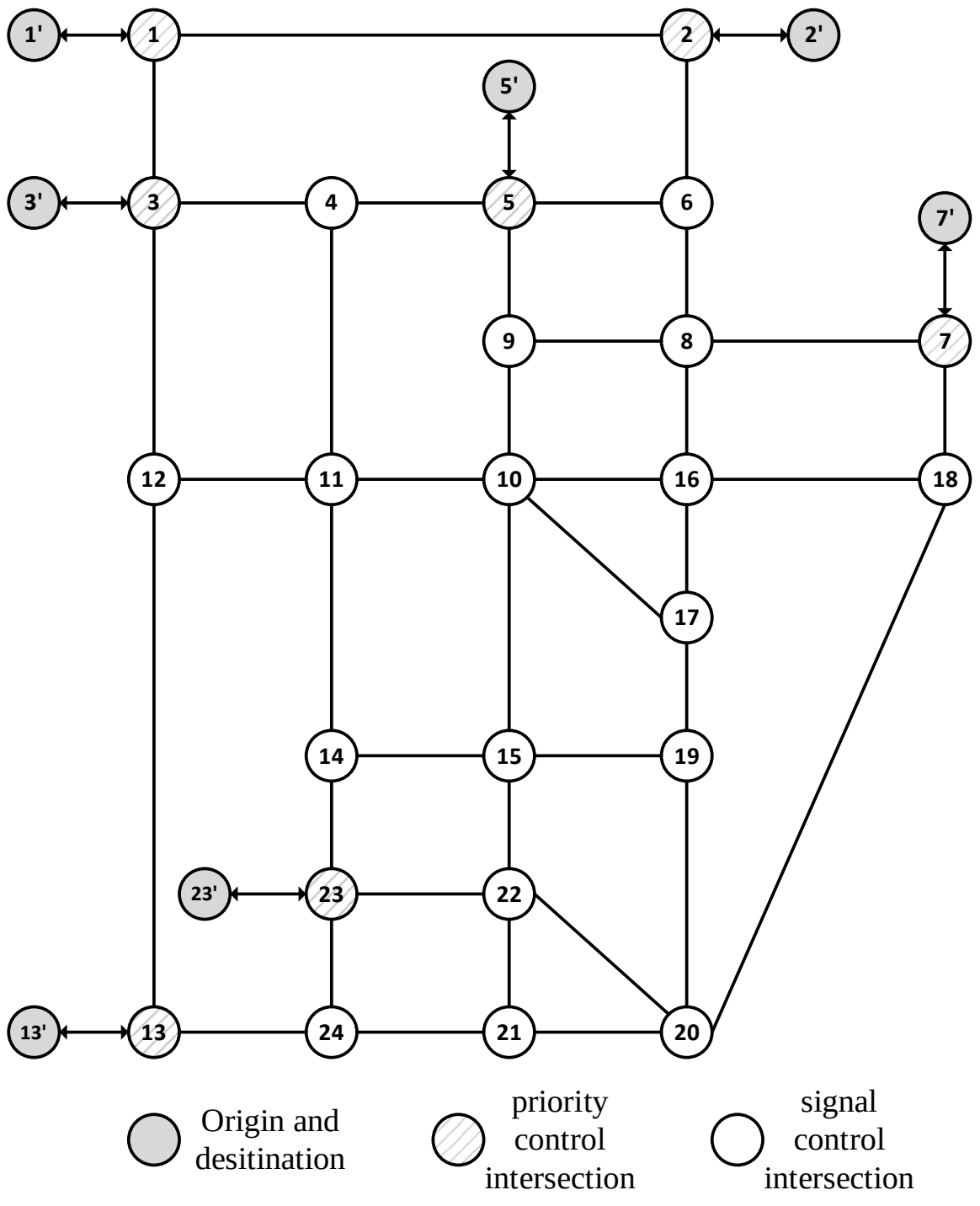

Figure 6 The modified Sioux Falls network

The edges in the road network are bidirectional with two lanes in each direction. As shown in Fig.6, there are three types of signalized intersections, respectively with three, four, five inbound roads. The layout and potential signal phases of these intersections are presented in Table B.1.

Time-varying traffic flows are applied to train the control policy and improve its efficiency under varying traffic conditions. Four types of flow distribution are designed and presented in Figure B.1. Flow ratio represents the proportion between the current flow and the peak flow. The OD pairs and their potential routes, peak flow and the type of flow fluctuations and corresponding RAs are shown in Table

B.2. In summary, the simulation environment consists of 17 SAs and 12 RAs.

### 5.1.2. Parameters of the MAA2C algorithm

In each episode, there are 720 control steps. The training is terminated after 500000 control steps, which is approximately 695 episodes. Other parameters necessary to implement the proposed MADRL algorithm are listed in Table B.3.

Within the framework of DRL, different indicators (such as different types of states, reward) may show different dimensions and unit, thereby impeding the accuracy of training outcomes. To counteract this, normalization techniques are employed to remove the influence of dimensionality and to facilitate comparability among such indicators. In particular, the normalized states and rewards are clipped within specific ranges. Normalization factors for each indicator are determined based on default values obtained from a sample simulation.

The TensorFlow function 'tf.train.AdamOptimizer' is utilized with several parameters to regulate and fine-tune the optimization process. The learning rate determines the magnitude of parameter updates during training iterations. The exponential decay rates for the first and second moment estimates control the influence of past gradients on the current update. A small constant, epsilon, is added to the denominator of the update rule to prevent division by zero. Also, we employ the gradient clipping to control the norm of gradients and mitigate the issue of exploding gradients. When the gradient norm surpasses a predefined threshold, the gradients are rescaled to keep them within the designated limit. This technique serves to stabilize the training process and prevent drastic parameter updates caused by excessively large gradients.

## 5.2. Training and Evaluation Results

Fig.8 illustrates the training curve for joint control, depicting the cumulative sum of local rewards per training episode across all agents, which can be expressed as follows:

$$R = \sum_{t \in T} \sum_{i \in C_A} r_{i,t}^{\text{local}} \tag{17.}$$

Where $T$ is the set of control steps per episode. The total training curve initially increases and then converges as MADRL algorithm learn from accumulated experience, ultimately reaching an optimal policy.

Fig.9 displays the cumulative local reward per episode for each signal control agent and vehicle routing agent. Initially, there are substantial fluctuations in the cumulative reward as the training progresses. However, these fluctuations gradually diminish over time, leading to a more stable reward trend. Notably, the rewards of most agents show considerable improvements throughout the training process.

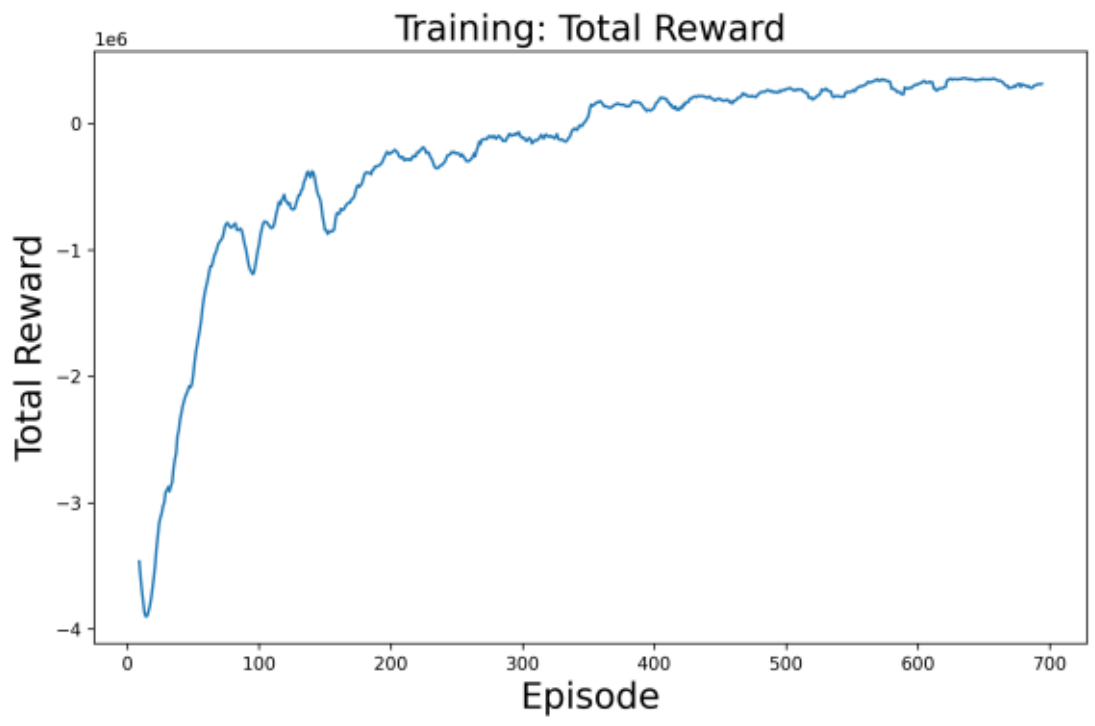

Figure 8 Training curves for joint control

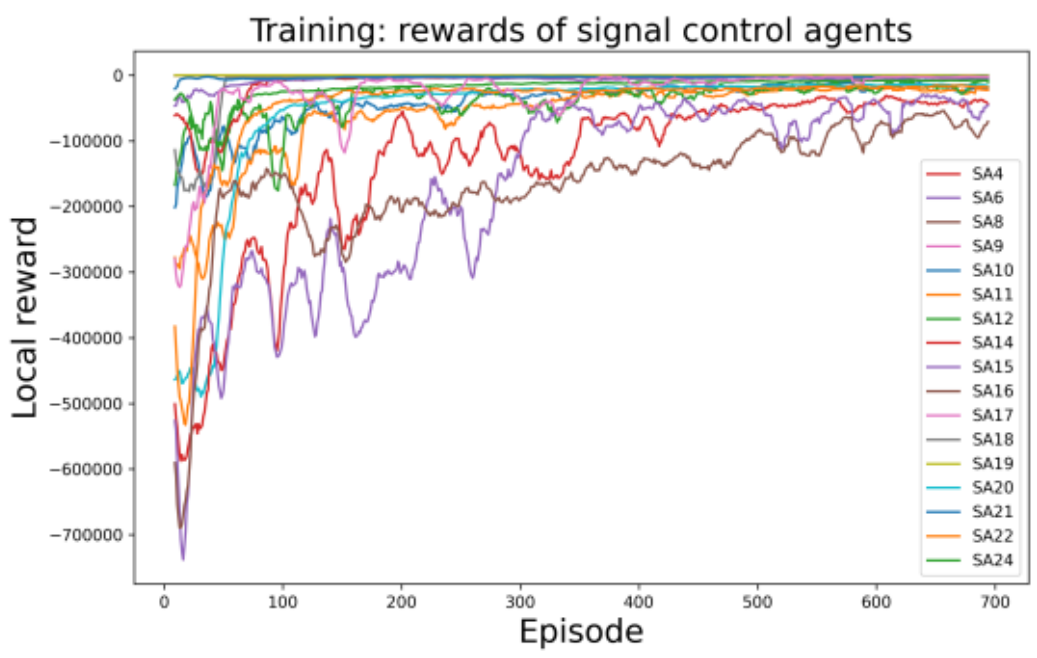

(a) Signal control agent

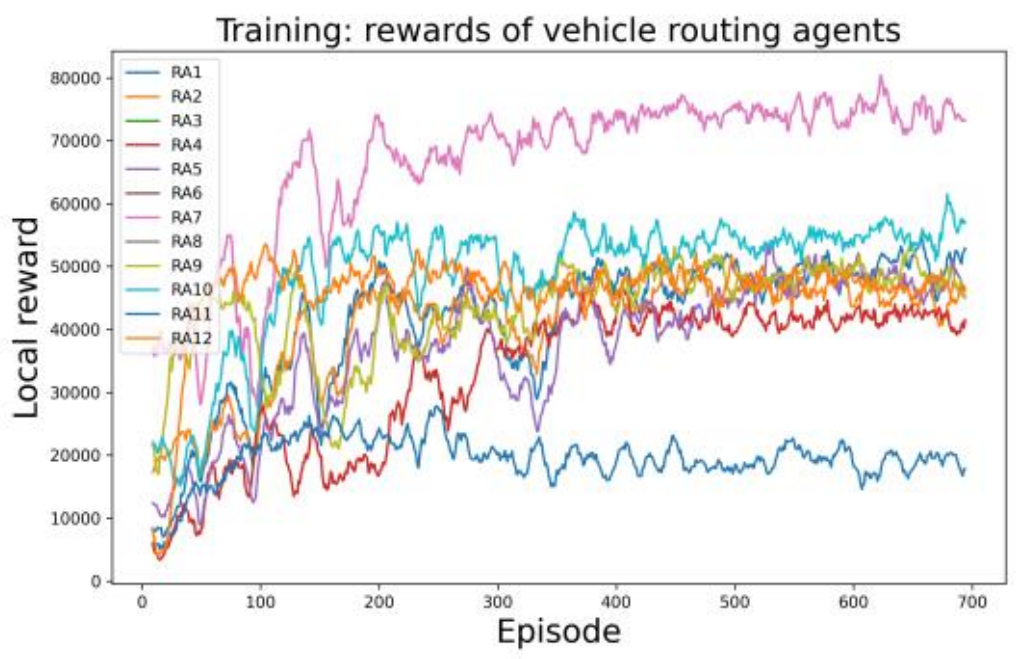

(b) Vehicle routing agent

Figure 9 Training curves for each agent

To provide a comprehensive understanding of the impact of joint optimization of signal control and vehicle routing, we also explore two control scenarios for comparison. The first scenario involves signal control without controlling the vehicles' routes, while the second scenario involves route guidance with fixed signal control. These scenarios are considered to further analyze and contrast the effects of different control strategies. Fig.10 illustrates the curves depicting the total number of arrived vehicles and the average vehicle delay during the training process for each scenario.

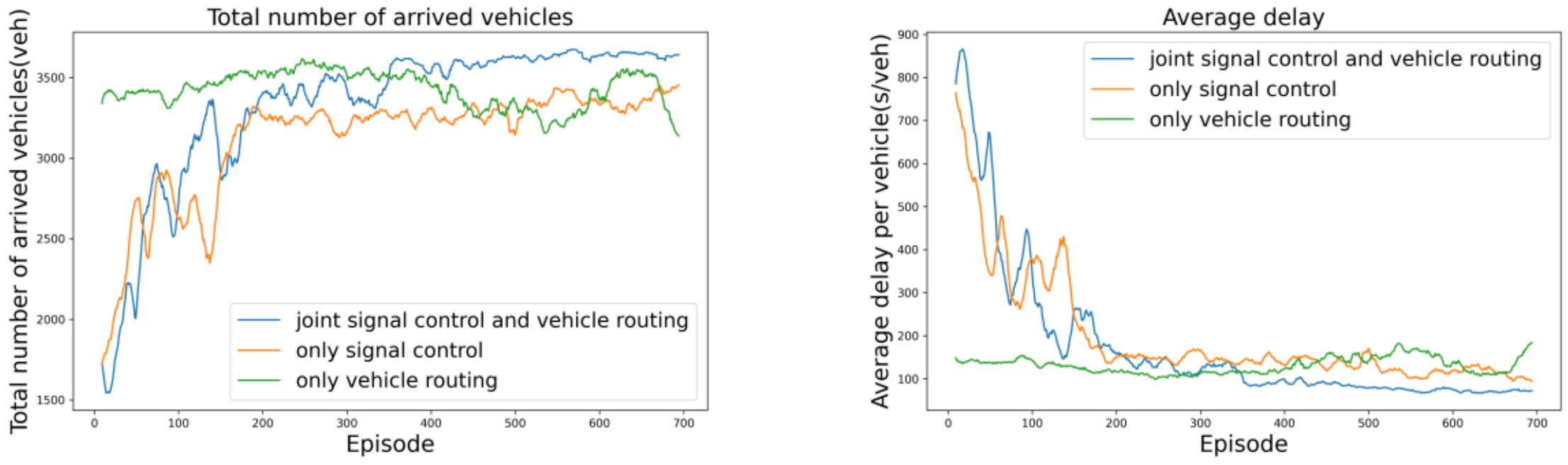

(a) Total number of arrived vehicles (b) Average vehicle delay

Figure 10 Training curves for three scenarios

According to Fig.10, the training curves for signal control and joint control show similar performance, with both converging within a few episodes. However, joint control surpasses signal control by establishing an enhanced control policy for both route and signal timings. This improvement leads to notable enhancements in network performance, including a higher number of arrived vehicles and a reduced average vehicle delay. On the other hand, the initial performance of the vehicle routing control is relatively high, owing to the initial random exploration and optimistic parameter settings that prompt the agents to take actions with higher rewards. However, as the training progresses, the agents fail to improve their strategies.

The optimized policy was implemented in the environment, running for 10 episodes to collect evaluation results for further analysis of the effectiveness of joint control. The evaluation metrics include the total number of arrived vehicles, average delay, average speed, and average queue per intersection approach. As shown in Table 2, it is evident that the joint control surpasses the individual approaches of signal control or vehicle routing, demonstrating superior performance.

Table 2 Evaluation results for three scenarios

| | Total arrived number (veh) | Average delay (s) | Average speed (m/s) | Average queue (veh) |
|---|---|---|---|---|
| Joint control | **3646** | **60.3** | **11.2** | **0.26** |
| Signal control | 3412 | 67.7 | 10.5 | 0.43 |
| Vehicle routing | 3157 | 86.6 | 8.6 | 0.73 |

### 5.3. Different Compliance

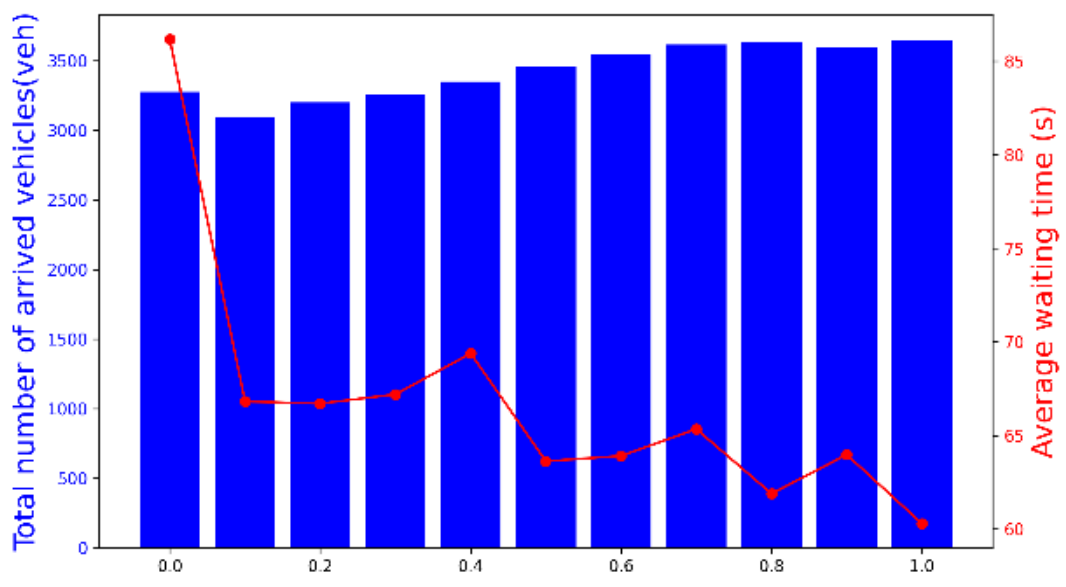

Figure 11 network performance under different vehicle compliance rates

We conducted an analysis of the optimized joint control policy's performance under varying vehicle compliance conditions. In our evaluation environment, we modeled traffic flows with two types of vehicles: those that comply with the system instructions and others that do not. The compliant vehicles follow the routes optimized by the control policy, while the non-compliant vehicles choose the pre-defined route. Fig.11 illustrates the network's performance under different compliance rates. The results suggest that higher compliance levels have a positive impact on the control effectiveness of the joint control policy. Specifically, the strategy performs well when the vehicle compliance rate exceeds 70%.

The above conclusion differs from that of many other methods. For example, in the study by Le et al. (2017), their proposed control strategy achieved optimal performance at a compliance rate of 30%. This was because their route guidance approach aims to divert traffic away from congested areas, thereby improving the overall travel time in the network. However, when too many vehicles follow the route guidance advice (more than 30%), the benefits of reduced congestion are outweighed by the increased travel time on longer paths, resulting in a decrease in overall network performance. This drawback highlights the limitations of distributed control, where route guidance advice is solely based on local information.

In our proposed method, vehicle routing relies on network-wide information, thus avoiding such limitations. Therefore, in our study, higher compliance with the optimized routes leads to better outcomes compared to lower compliance levels. This finding provides further evidence supporting the superiority of our control policy solution.

### 5.4. Different Traffic Condition

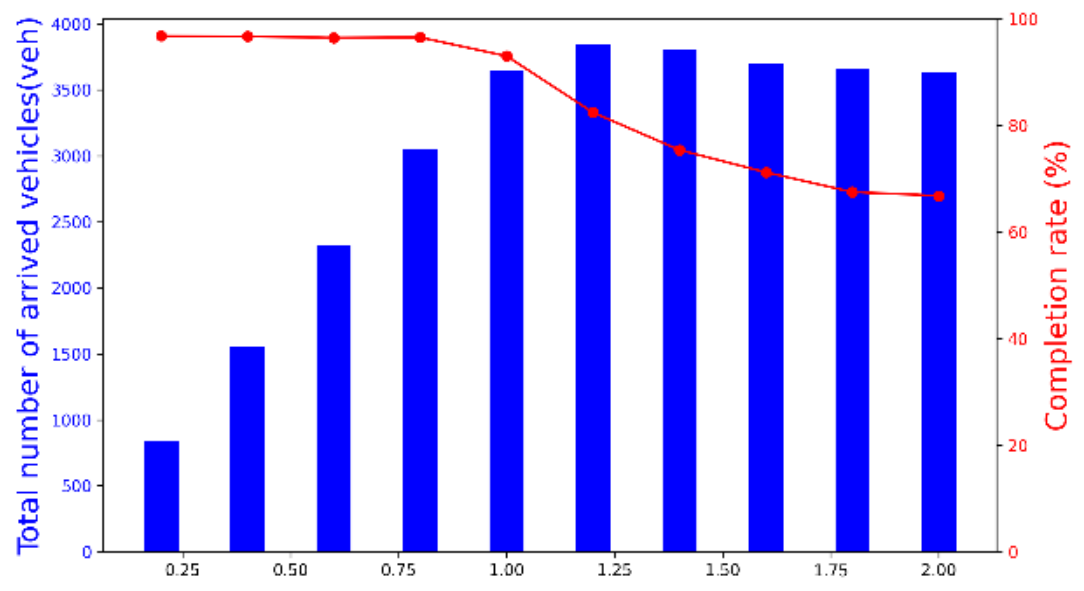

Figure 12 Performance under different traffic condition

Fig.12 illustrates the performance of the optimized control policy under varying traffic conditions. The x-axis represents the multiplicative factor of the current traffic volume, the bars represent the total number of arrived vehicles, and the line represents the completion rate, which is calculated by dividing the number of arrived vehicles by the number of departed vehicles.

When the multiplicative factor is below 1.25, the throughputs of the situations increase as the traffic demand increases. However, when the multiplicative factor exceeds 1.25, this trend reverses, and a decrease in throughputs is observed. Additionally, the completion rate reaches its highest level when the multiplicative factor is below 0.8. This indicates that at these levels, the traffic volume is undersaturated, and all demands can be promptly met.

## 6. Conclusion

The joint optimization approach proposed in this paper, which combines traffic signal control and vehicle routing, has shown promising results in improving network performance and reducing traffic congestion. By employing MAA2C algorithm, the approach enables signal control agents to optimize signal timings and vehicle routing agents to optimize route choice. Through defining relevance among agents and enabling them to share observations and rewards, the proposed method promotes interaction and cooperation among agents. Additionally, the DNN structure is designed to facilitate algorithm convergence. To our knowledge, this study is the first to employ MADRL to find the optimal joint policy of signal control and vehicle routing.

The numerical experiments conducted on the modified Sioux network have demonstrated that the proposed joint signal control and vehicle routing surpasses the individual strategies of controlling signal timings alone or controlling vehicles' routes alone. This integrated approach proves to be more effective in enhancing the total number of arrived vehicles and average speed, while simultaneously reducing average delay and queue length. Furthermore, the study explores the performance of the optimized joint control policy across various vehicle compliance conditions.

In conclusion, the joint optimization approach proposed in this paper has proven to be effective in reducing delays and improving throughput in signalized road networks. The approach can also be extended to other vehicle-road collaboration problems, such as joint variable speed limits and ramp control, as well as joint signal control and vehicle speed control. However, it is important to acknowledge that the approach has certain limitations. One drawback is the potential increase in computational complexity and cost, particularly for large-scale networks, as the MADRL method requires significant computational resources for agent training. In addition, the cooperation parameters that may impact the performance of the approach must be accurately defined before training. Future research can incorporate this process into training by using Graph Neural Network (GNN) to estimate the value and policy networks as well as to illustrate the cooperation among agents. Overall, the proposed approach demonstrates the potential of MADRL in tackling intricate traffic management challenges and offers valuable perspectives for future studies.

**Appendix A**

Table A.1 Details of the relevant studies

| Study | Control mode | Control strategy | Objective | Criterion for route | Decision variable | Travel cost estimation | Algorithm |
|---|---|---|---|---|---|---|---|
| (Chiou 2014) | Static traffic | Centralized control | Minimize delay and number of stops | User equilibrium | Signal settings variables; link flows; OD demand growth factors | TRANSYT | Bundle-type gradient method |
| (Sheffi and Powell 1983) | | | Minimize total travel time | | Green split | Link volume-travel time curve | Gradient projection algorithm |
| (Li, | Dynami | | | System | Phase-time arc | Space- phase-time | Lagrangian |
| Mirchandani, and Zhou 2015, Wang et al. 2020) | c traffic | | | optimization | | hyper network | decomposition method |
| (Varia and Dhingra 2004) | | | | | Green time; flow distribution | Delay model for signalized intersection | Genetic algorithm |
| Yu, Ma, and Zhang (2018) | | | | User equilibrium | Green time | Continuous-time double-queue model | |
| (Han et al. 2015) | | | Minimize network-wide travel costs | | Green splits | LWR-based dynamic network | Simulated annealing method, |
| | | | | | | model | etc |
| (Hawas 2000) | | Distributed control | Minimize total travel time | System optimization | Green time, guided routes | A microscopic simulation model | On-line operation |
| (Le et al. 2017) | | | Maximize network throughput and minimize spatial heterogeneity | | Signal phase, turning fractions | - | BackPressure mechanism's principle |

## Appendix B

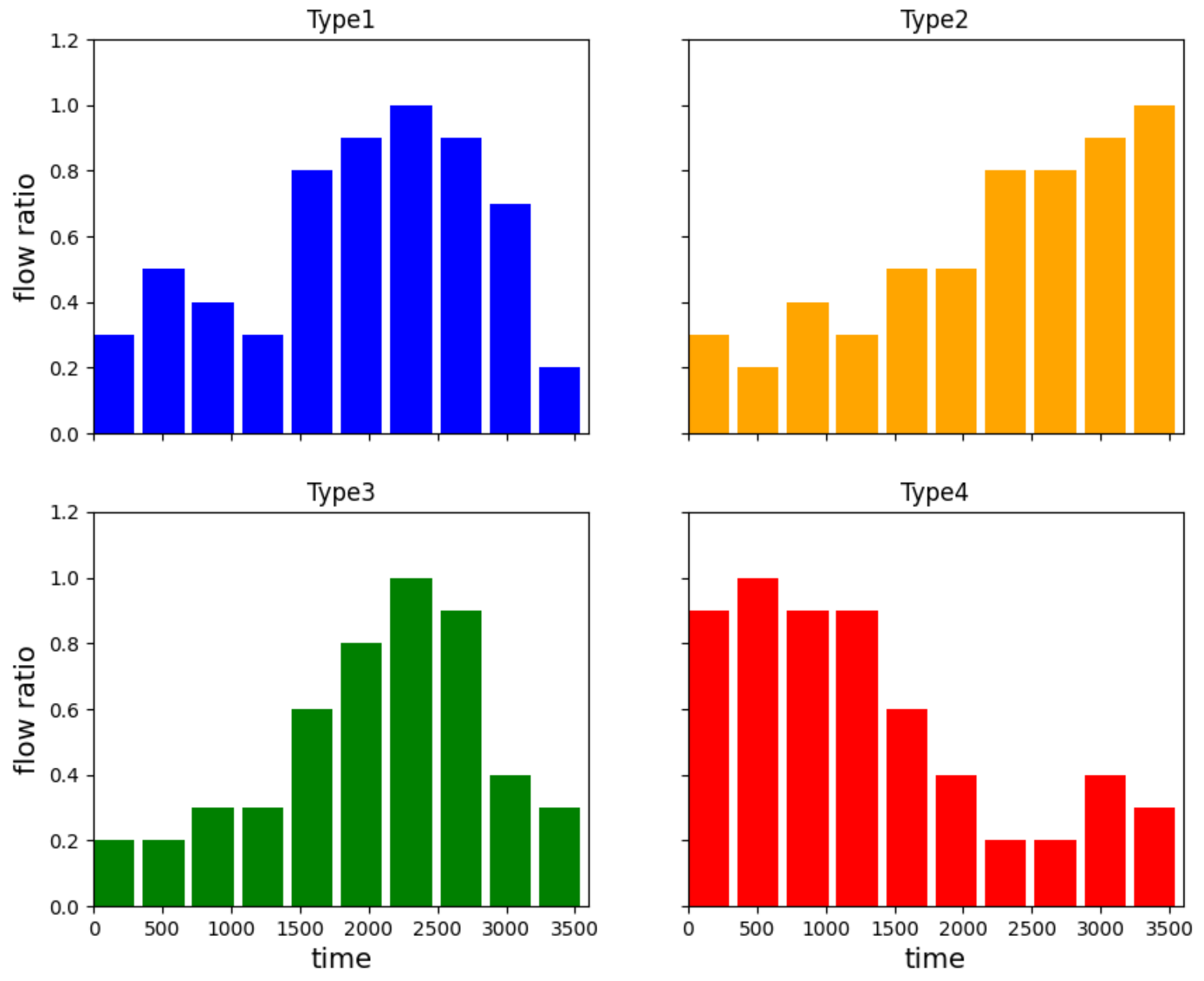

Figure B.1 Four types of flow distribution

Table B.1 The layout and potential signal phases of intersections in the network

| Type | Layout | Phase pattern |
|---|---|---|
| 1 | 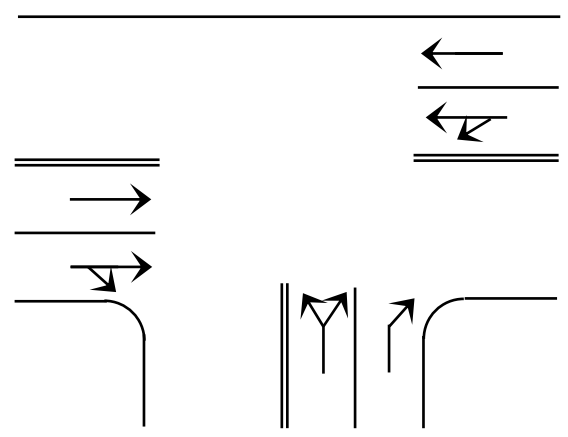 | 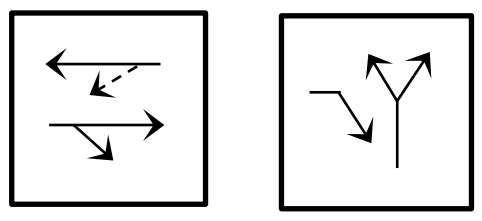 |
| 2 | 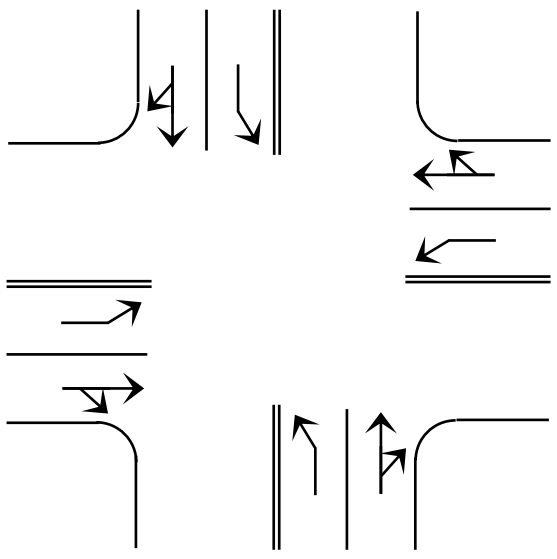 | 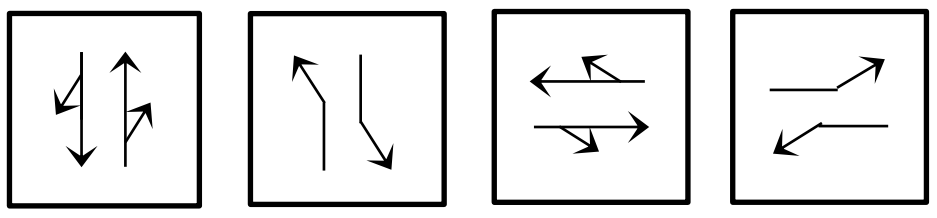 |
| 3 | 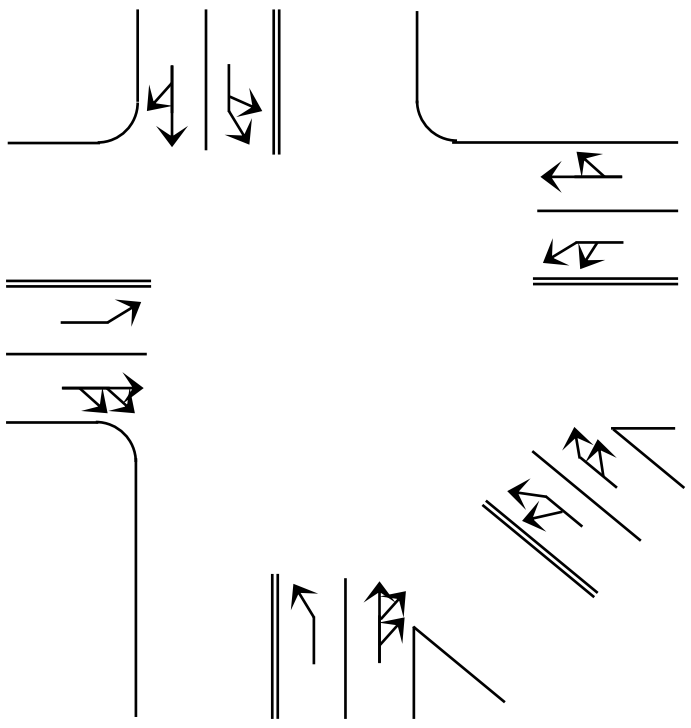 | 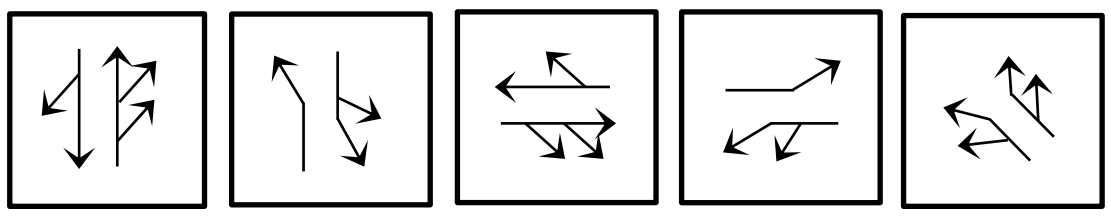 |

Table B.2 Traffic demand

| Index | Origin | Destination | Peak flow | Fluctuations type | Potential routes | RA |
|---|---|---|---|---|---|---|
| 1 | 1 | 2 | 240 | 1 | [1 2] | |
| 2 | 1 | 5 | 240 | 2 | [1 3 4 5] | |
| 3 | 1 | 7 | 360 | 3 | [1 3 4 5 6 8 7], [1 3 4 5 9 8 7] | RA1 |
| 4 | 1 | 23 | 360 | 2 | [1 3 4 11 14 23], [1 3 12 11 14 23] | RA2 |
| 5 | 2 | 1 | 120 | 1 | [2 1] | |
| 6 | 2 | 5 | 120 | 3 | [2 6 5] | |
| 7 | 2 | 7 | 240 | 4 | [2 6 8 7] | |
| 8 | 2 | 13 | 240 | 1 | [2 1 3 12 13] | |
| 9 | 2 | 23 | 360 | 3 | [2 6 8 16 10 15 22 23],<br>[2 6 5 4 11 14 23],<br>[2 6 8 16 10 11 14 23] | RA3,<br>RA4 |
| 10 | 3 | 1 | 120 | 2 | [3 1] | |
| 11 | 3 | 7 | 360 | 3 | [3 4 5 6 8 7],<br>[3 4 5 9 8 7] | RA5 |
| 12 | 5 | 2 | 240 | 4 | [5 6 2] | |
| 13 | 5 | 23 | 480 | 1 | [5 9 10 15 22 23],<br>[5 4 11 14 23],<br>[5 9 10 11 14 23] | RA6,<br>RA7 |
| 14 | 7 | 1 | 360 | 2 | [7 8 6 2 1],<br>[7 8 6 5 4 3 1],<br>[7 8 9 5 4 3 1] | RA8,<br>RA9 |
| 15 | 7 | 2 | 120 | 3 | [7 8 6 2] | |
| 16 | 7 | 13 | 240 | 4 | [7 18 20 21 24 13] | |
| 17 | 7 | 23 | 240 | 2 | [7 18 20 22 23] | |
| 18 | 13 | 2 | 360 | 1 | [13 24 23 14 11 4 5 6 2],<br>[13 12 3 1 2] | RA10 |
| 19 | 13 | 3 | 240 | 3 | [13 12 3] | |
| 20 | 13 | 5 | 360 | 4 | [13 12 3 4 5] | |
| 21 | 13 | 7 | 240 | 1 | [13 24 21 20 18 7] | |
| 22 | 13 | 23 | 240 | 2 | [13 24 23] | |
| 23 | 23 | 1 | 180 | 3 | [23 14 11 4 3 1],<br>[23 14 11 12 3 1] | RA11 |
| 24 | 23 | 5 | 120 | 2 | [23 14 11 4 5] | |
| 25 | 23 | 7 | 360 | 4 | [23 22 20 18 7],<br>[23 22 15 10 9 8 7] | RA12 |
| 26 | 23 | 13 | 240 | 1 | [23 24 13] | |

Table B.3 Hyperparameters for MADRL training

| Description | | | | | Applied value |
|---|---|---|---|---|---|
| The number of steps in a batch $n$ | | | | | 144 |
| Discount factor $\gamma$ | | | | | 0.99 |
| Entropy coefficient of regularization term $\beta_{\text{en}}$ for SA | | | | | 0.05 |
| Entropy coefficient of regularization term $\beta_{\text{en}}$ for RA | | | | | 0.01 |
| The threshold for the gradient norm | | | | | 30 |
| Adam optimizer | Learning rate | | | | 2.5e-4 |
| | Exponential decay rates for the first estimates | | | | 0.9 |
| | Exponential decay rates for the second estimates | | | | 0.999 |
| | Epsilon | | | | 1e-8 |
| DNN setting | Neuron number of LSTM | | | | 128 |
| | Neuron number of FC1 | SA | 140 | RA | 20 |
| | Neuron number of FC2 | | 60 | | 70 |
| | Neuron number of FC3 | | 50 | | 200 |
| | Neuron number of FC4 | | 70 | | 50 |
| State clip | Clipping threshold | | | | [0,2] |
| | the normalization factors of state $s_{i,t,l}^{\text{wave}}$ | | | | 2 |
| | the normalization factors of state $s_{i,t.l}^{\text{wait}}$ | | | | 8 |
| | the normalization factors of state $s_{i,t,e}^{\text{arrival}}$ | | | | 16 |
| | the normalization factors of state $s_{i,t,e}^{\text{down}}$ | | | | 7 |
| Reward clip | Clipping threshold | | | | [-6,6] |
| | the normalization factors of SA's reward $r_{i,t}$ | | | | 400 |
| | the normalization factors of RA's reward $r_{i,t}$ | | | | 100 |
| Coefficient of reward, state | $\alpha_1 = 1,\ \alpha_2 = 1000,\ \beta^{SS} = 0.3,\ \beta^{SR} = 0.3,\ \beta^{RS} = 0.1,\ \delta = 0.5,\ \sigma = 0.5$ | | | | |